\documentstyle[12pt,epsfig,rotating]{article}
\def\pge{\pagestyle{empty}} \def\pgn{\pagestyle{plain}}
\def\bsls35{\baselineskip 0.35in}
     \def\d{{\rm d}}
\def\spg{\setcounter{page}} 
\def\bd{
\begin{document}} \def\ed{\end{document}}
\def\bmp{\begin{minipage}} \def\emp{\end{minipage}}
\def\bcc{\begin{center}} \def\ecc{\end{center}}     \def\npg{\newpage}
\def\beq{\begin{equation}} \def\eeq{\end{equation}} \def\hph{\hphantom}
\def\be{\begin{equation}} \def\ee{\end{equation}} \def\r#1{$^{[#1]}$}
\def\n{\noindent} \def\ni{\noindent} \def\pa{\parindent} 
\def\hs{\hskip} \def\vs{\vskip} \def\hf{\hfill} \def\ej{\vfill\eject} 
\def\cl{\centerline} \def\ob{\obeylines}  \def\ls{\leftskip}
\def\underbar#1{$\setbox0=\hbox{#1} \dp0=1.5pt \mathsurround=0pt
   \underline{\box0}$}   \def\ub{\underbar}    \def\ul{\underline} 
\def\f{\left} \def\g{\right} \def\e{{\rm e}} \def\o{\over} 
\def\vf{\varphi} \def\pl{\partial} \def\cov{{\rm cov}} \def\ch{{\rm ch}}
\def\la{\langle} \def\ra{\rangle} \def\EE{e$^+$e$^-$}
\def\bitz{\begin{itemize}} \def\eitz{\end{itemize}}
\def\btbl{\begin{tabular}} \def\etbl{\end{tabular}}
\def\btbb{\begin{tabbing}} \def\etbb{\end{tabbing}}
\def\beqar{\begin{eqnarray}} \def\eeqar{\end{eqnarray}}
\def\\{\hfill\break} \def\dit{\item{-}} \def\i{\item} 
\def\bbb{\begin{thebibliography}{9} \itemsep=-1mm}
\def\ebb{\end{thebibliography}} \def\bb{\bibitem}
\def\bpic{\begin{picture}(260,240)} \def\epic{\end{picture}}
\def\akgt{\noindent{\bf Acknowledgements}}
\def\fgn{\noindent{\bf\large\bf Figure captions}}
\bd
\pge

\null{}\vskip -1.2cm
\hskip12cm{\bf HZPP-0001}
\vskip-0.2cm

\hskip12cm Jan. 20, 2000

\vskip1cm

\begin{center}
{\Large The Influence of Multiplicity Fluctuation
\vskip0.4cm

on the Erraticity Behaviour in High Energy Collisions\footnote{This 
work is supported in part by the NSFC under project 19975021.}}
\vskip0.5cm

\vskip0.2cm

{\large Liu zhixiu, \ \  Fu Jinghua \ \ and \ \ Liu Lianshou}

{\small Institute of Particle Physics, Huazhong Normal University,
Wuhan 430079 China}
\date{ }
\end{center}

\begin{center}
\begin{minipage}{125mm}
\vskip 0.5in
\begin{center}{\Large ABSTRACT}\end{center}
{\hskip0.6cm
The influence of multiplicity distribution (fluctuation of multiplicity
in the event space) on the erraticity behaviour
in high energy collisions is investigated via Monte Carlo simulation
and compared with the experimental results from NA27 data. It is shown
that, the erraticity phenomenon is insensitive to the multiplicity
fluctuation. When the average multiplicity is low, this phenomenon
is dominated by the statistical fluctuations also in the case of
multiplicities, fluctuating from event to event. }
\end{minipage}
\end{center}
\vs0.8cm
{\large PACS number: 13.85 Hd

\ni
}

\npg \pgn \spg{2}
\baselineskip 0.24in

Since the discovery of unexpectedly large local fluctuations
in a high multiplicity event recorded by the JACEE
collaboration~\cite{JACEE}, the investigation of non-linear phenomena in
high energy collisions has attracted much attention~\cite{Kittel}. An
anomalous scaling of factorial moments, defined as
\beqar 
F_q &=&
    \frac{1}{M}\sum\limits_{m=1}^{M}
    \frac{\la n_m(n_m-1) \cdots (n_m-q+1)\ra}
    {\la n_m \ra ^q  } ,
\eeqar
at diminishing phase space scale or increasing partition number
$M$ of phase space~\cite{BP}:
\beqar 
F_q \propto M^{-\phi_q},
\eeqar
called intermittency (or fractal) has been proposed for this purpose
in multiparticle system and has been observed successfully in various 
experiments~\cite{NA22}\cite{NA27}. The average $\la \cdots \ra$ in 
Eqn.(1) is over the whole event sample and $n_m$ is the number of 
particle falling into the $m$th bin.  

Recently, Cao and  Hwa~\cite{CaoHwa} argued that besides the factorial 
moments $F_q$ averaged over event sample, the fluctuation of single event 
factorial moments  
\beqar  
F_q^{({\rm e})} &=& \frac
    { \frac{1}{M}\sum\limits_{m=1}^{M} n_m(n_m-1) \cdots (n_m-q+1)}
    { \f( \frac{1}{M}\sum\limits_{m=1}^{M} n_m\g)^q}
\eeqar 
in the event space can also be ultilized to characterize the dynamical
property of the hadronic system produced in the collisions.

The fluctuations of $F_q^{\rm e}$  from event to event can be quantified
by its normalized moments as:
\begin{equation} 
C_{p,q}=\la \Phi_q^p\ra, \quad \Phi_q= {F_q^{(e)}} \f/
\la F_q^{(e)}\ra \g.,
\end{equation}
\noindent and by $\d C_{p,q}/\d p$ at $p=1$:
\beqar   
\Sigma_q=\la \Phi_q \ln \Phi_q \ra
\eeqar
If there is a power-law behavior of the fluctuation as the partition
number goes to infinity (or as the resolution $\delta=\Delta/M$
becomes very small), {\rm i.e.}
\begin{equation}   
C_{p,q}(M) \propto M^{\psi_q(p)},
\end{equation}
the phenomenon is referred to as erraticity~\cite{hwa}.
The derivative of the exponent $\psi_q(p)$ at $p=1$
\begin{equation}   
\mu_q=\f.\frac{\d}{\d p}\psi_q(p)\g|_{p=1} =
             \frac{\partial \Sigma_q}{\partial \ln M}
\end{equation}
describes the width of the fluctuation and is called the entropy index. 

Wang et al.~\cite{WSS} have used this method to analyse the 400 GeV/$c$ 
pp collision data from NA27 and really found evidences for the predicted 
erraticity phenomenon. Some physical conclusions have been drawn tentatively 
from this experimental finding.

However, it is well known that the obstacle of event-by-event analysis is
the influence of statistical fluctuations caused by an insufficient number
of particles.  The big advantage~\cite{BP} of sample factorial moments, 
Eqn.(1), in being able to eliminate the statistical fluctuations comes from
the average over event sample. The same procedure cannot be applied to
the event factorial moments of Eqn.(3)~\cite{PLB}\cite{zgkx}. Since the 
number of particles in an event is always finite, event
factorial moments cannot completely eliminate statistical fluctuations
and therefore cannot represent the dynamical probability
moments associated with it. 

It has been shown~\cite{zgkx} that a flat probability distribution with 9 
particles in each event, distributed according to the Bernoulli distribution:
\beq   
B(n_1,\dots,n_M|p_1,\dots,p_M) = \frac{N!}{n_1!\cdots n_M!}
p_1^{n_1}\cdots p_M^{n_M} , \qquad \sum_{m=1}^M n_m=N,
\eeq
can reproduce the phenomenon observed in NA27 data~\cite{WSS}. This indicates 
that when multiplicity is low the fluctuation of factorial moments in event
space is dominated by statistical fluctuations~\cite{PLB}. 
The erraticity phenomenon observed in NA27 data is mainly due to statistical
fluctuations and has little to do with real physics.

However,
in order to draw a more reliable conclusion, it should be noticed that
the multiplicity in NA27 experiment is not a constant, but is fluctuating
from event to event. How will this
additional fluctuation in the event space influence the fluctuation
in the same space of the factorial moment? To answer to this question 
is the main goal of this letter.

Let us parametrize the multiplicity distribution in 400 GeV/$c$ pp collision
from NA27 experiment~\cite{WSSpn} by means of a negative-binomail 
distribution (NBD)~\cite{NBD}:
\beq  
P_n = {n+k-1 \choose n} \f(\frac{\la n\ra/k}{1+\la n\ra/k}\g)^n
   \frac{1}{(1+\la n\ra/k)^k}.
\eeq
In Eqn.(9) the average multiplicity $\la n\ra$ is taken as 9.84, the 
parameter $k$ is fitting to be 12.76, cf. Fig.1.

According to the NBD distribution, Eqn.(9), 60 000 events are obtained 
using Monte Carlo method.  In each event the phase space region $\Omega$
is divided into $M$ bins and the $n$ particles of this event are allocated 
in these bins according to the Bernoulli distribution, Eqn.(8). The event 
factorial moments $F_q^{({\rm e})}$ and the corresponding 
$C_{p,q}$, $\Sigma_q$ are then calculated according to Eqn's.(3), (4), (5).

The results of ln$C_{p,q}$ versus ln$M$ ($q=2,3,4$) are plotted in Fig.2
together with the experimental results~\cite{WSS} from NA27 data. 
Comparing the results of the present
model with both statistical and multiplicity fluctuations
with the results in Ref.~\cite{zgkx} for fixed 
multiplicity ($n=9$), having only statistical 
fluctuations, we can see that the influence of multiplicity fluctuation
is weak.  When the average multiplicity is low, the erraticity phenomenon
is still dominated by the statistical fluctuations, even in the case of
multiplicities, fluctuating from event to event. 

In order to show the dependence of erraticity behaviour on the width of
multiplicity distribution, i.e. the strength of multiplicity fluctuation 
in the event space, in more detail, we have done the same calculation by 
fixing the average multiplicity to $\la n\ra=9$ while varying the 
distribution  width through changing the NBD parameter 
$k$ from 0.5 to 18, cf. Fig.3. The results for
ln$C_{p,q}$ and $\Sigma$ are shown in Fig's 4($a$) and ($b$) respectively.
It can be seen from Fig's.3 and Fig.4 that, when the parameter $k$ 
varies within the above-mentioned  range the width of multiplicity 
distribution changes considerably while the 
erraticity phenomenon remains qualitatively the same.
The corresponding curves are nearly parallel to each other at large $M$.

In conlusion, we have been able to reproduce the erraticity phenomenon 
observed in the experimental data from NA27, using a flat probability 
distribution with both statistical and multiplicity fluctuations. 
This means that when the average multiplicity is low, the erraticity 
phenomenon is dominated by the statistical fluctuations also in the case 
of multiplicities, fluctuating from event to event.  Through varying 
the NBD parameter $k$ over a large range, we have shown that
the erraticity phenomenon is insensitive to the width of multiplicity
distribution.

\vskip0.5cm
\n
Acknowledgement \ The authors are grateful to Wu Yuanfang
for valuable discussions.

\newpage
\baselineskip 0.36in
\def\Journal#1#2#3#4{{#1} {\bf #2} (#3) #4}
\def\NCA{\em Nuovo Cimento} \def\NIM{\em Nucl. Instrum. Methods}
\def\NIMA{{\em Nucl. Instrum. Methods} A} \def\NPB{{\em Nucl. Phys.} B}
\def\PLB{{\em Phys. Lett.}  B} \def\PRL{\em Phys. Rev. Lett.}
\def\PRD{{\em Phys. Rev.} D} \def\ZPC{{\em Z. Phys.} C}
\def\PRE{{\em Phys. Rev.} E} \def\PRC{{\em Phys. Rev.} C}

\newpage
\n{\Large Figure captions}
\vs0.5cm
\n
Fig.1 \ The negative-binomial distribution ($\la n\ra=9.84$, $k = 12.76$) 
fitting to the multiplicity distribution 
of 400 GeV/$c$ pp collision. Data taken from \cite{WSSpn}.

\n
Fig.2 \ The $C_{p,q}$ of flat probability distribution with both 
statistical and multiplicity fluctuations as compared with that 
of 400 GeV/$c$ pp collision. Data taken from \cite{WSSpn}.

\n
Fig.3 \ The negative-binomial distribution with various values of
parameter $k$.

\n
Fig.4 \ The $C_{p,q}$ ($a$) and $\Sigma_q$ ($b$) from flat probability 
distribution with statistical fluctuation and negative-binomial multiplicity 
distribution (with various values of parameter $k$).

\newpage

\begin{picture}(250,450)
\put(-55,-240)
{\epsfig{file=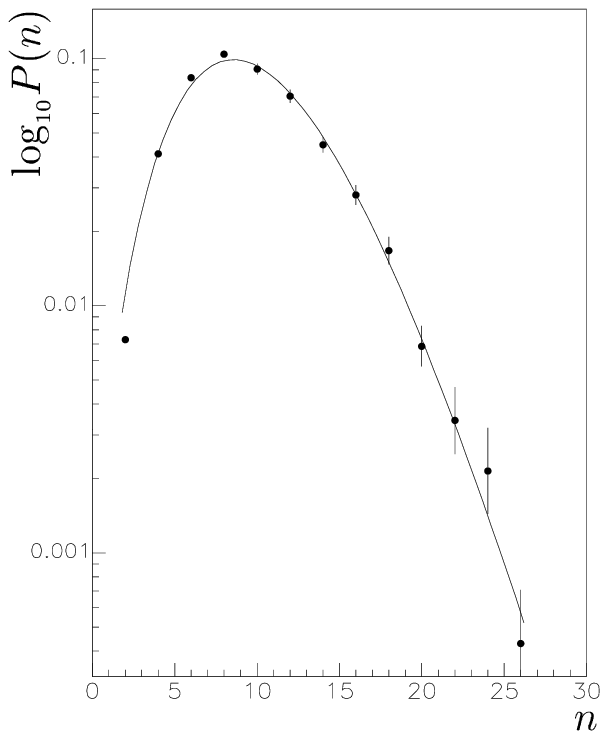,bbllx=0cm,bblly=0cm,
           bburx=8cm,bbury=6cm}}

\put(-90,-400)
{\epsfig{file=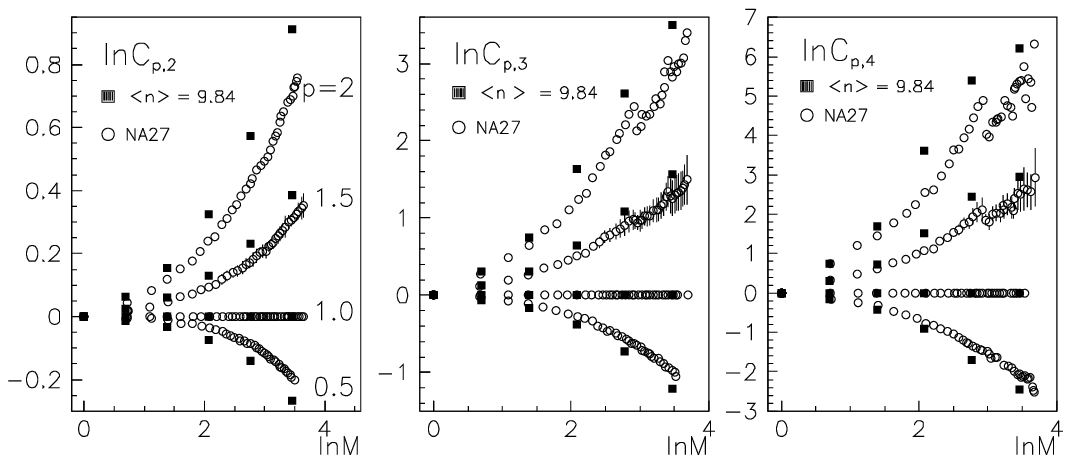,bbllx=0cm,bblly=0cm,
           bburx=8cm,bbury=6cm}}
 
\end{picture}

\vskip-8cm
\cl{Fig.1}
\vskip10cm
\cl{Fig.2}

\newpage

\begin{picture}(250,450)
\put(-75,-220)
{\epsfig{file=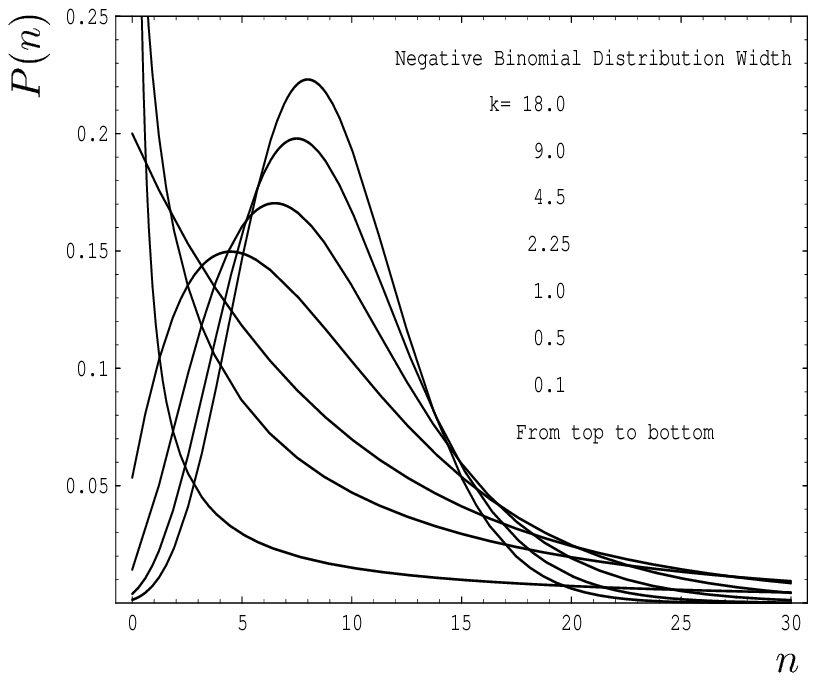,bbllx=0cm,bblly=0cm,
           bburx=8cm,bbury=6cm}}
\end{picture}
\vskip-10cm
\cl{Fig.3}

\begin{picture}(250,450)
\put(-95,-420)
{\epsfig{file=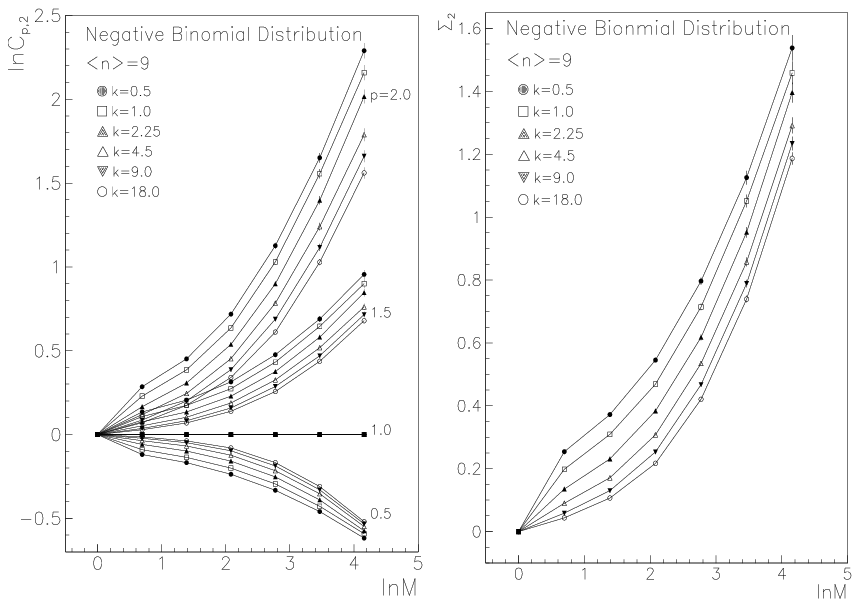,bbllx=0cm,bblly=0cm,
           bburx=8cm,bbury=6cm}}

\end{picture}

\vskip-6cm
\cl{($a$)\hskip5cm ($b$)}
\cl{Fig.4}

\ed